\documentclass[aip,jap,amsmath,floatfix,reprint]{revtex4-1}
\usepackage{graphicx,psfrag,mathptmx}

\begin{document}

\title{Structural evolution of Re (0001) thin films grown on Nb (110) surfaces by molecular beam epitaxy}
\author{Paul B. Welander}\email{welander@ll.mit.edu}
\affiliation{Lincoln Laboratory, Massachusetts Institute of Technology, Lexington, MA 02420}
\date{November 19, 2010}
\begin{abstract}
The heteroepitaxial growth of Re~(0001) films on Nb~(110) surfaces has been investigated.  Nb/Re bilayers were grown on A-plane sapphire -- $\alpha$-Al$_2$O$_3$~$(11\bar{2}0)$ -- by molecular beam epitaxy.  While Re grew with a (0001) surface, the in-plane epitaxial relationship with the underlying Nb could be best described as a combination of Kurdjumov-Sachs and Nishiyama-Wassermann orientations.  This relationship was true regardless of Re film thickness.  However, an evolution of the surface morphology with increasing Re thickness was observed, indicative of a Stranski-Krastanov growth mode.  Re~(0001) layers less than 150~{\AA} thick were atomically smooth, with a typical rms roughness of less than 5~{\AA}, while thicker films showed granular surface structures.  And despite the presence of a substantial lattice misfit, the Re layer strain diminished rapidly and the Re lattice was fully relaxed by about 200~{\AA}.  The strain-free and atomically smooth surface of thin Re overlayers on Nb is ideal for the subsequent epitaxial growth of ultra-thin oxide tunnel barriers.  Utilizing bcc/hcp (or bcc/fcc) heteroepitaxial pairs in advanced multi-layer stacks may enable the growth of all-epitaxial superconductor/insulator/superconductor trilayers for Josephson junction-based devices and circuits.
\end{abstract}
\maketitle

\section{Introduction}

Superconducting quantum bits fabricated from epitaxial Al$_2$O$_3$ tunnel barriers grown on Re (0001) have been shown to exhibit fewer two-level fluctuators than amorphous barriers formed by the \textit{in situ} oxidation of Al metal.\cite{Oh2006a}  This finding stands in stark contrast with epitaxial Al$_2$O$_3$ barriers grown on Nb (110), which were found to be electrically leaky.\cite{Welander2007}  The superior tunneling behavior of ultra-thin Al$_2$O$_3$ films on Re has been ascribed to the following: 1.) Re (0001) is very well lattice-matched with C-plane sapphire, $\alpha$-Al$_2$O$_3$ (0001), with an isotropic misfit of only $0.4\%$,\cite{Liu1970,Newnham1962} whereas the misfit between C-plane sapphire and Nb (110) is as large as $20\%$ along the Nb [001] direction;\cite{Barns1968} and 2.) Re resists oxidation to a very high degree, whereas Nb is a well-known getter.\cite{Samsonov1982}  The drawback with epitaxial Re films, however, is that they do not grow atomically smooth, with rms roughness of about 20~{\AA} typical.\cite{Oh2006b}  This is likely due to twinning or stacking faults in the film, a common problem for close-packed metal epitaxy.  Nb films grown on sapphire, on the other hand, are single-crystal and atomically smooth, with rms roughness an order of magnitude lower and a surface morphology comprised of 1000~{\AA}-wide terraces separated by monolayer step-edges.\cite{Wolfing1999,WelanderThesis}

This Letter describes a new approach to address the issue of epitaxial Re surface roughness.  Relatively thick Nb (110) layers, with their atomically smooth surfaces, were utilized for the subsequent growth of thin Re (0001) films.  Thick Re layers grown on Nb showed surface morphologies similar to those observed for thick Re layers grown directly on sapphire.  However, if the overlayer was kept thin then the Re surface retained the smooth morphology of the underlying Nb.  In addition, measurements of surface lattice parameters during film growth showed that any misfit strain in the Re was quickly relaxed.  Diffraction measurements indicated domain growth with multiple in-plane orientations, suggesting an imperfect registry with the underlying Nb that may be due, in part, to mixing at the bilayer interface.  Despite the shortcomings, these results support the use of multiple materials in the growth of epitaxial base electrodes, combinations of which could yield an ideal surface for single-crystal tunnel barrier growth -- one that is atomically smooth, well lattice-matched, and resistive to oxidation or inter-layer diffusion.

\section{Sample Preparation \& Analysis}

For this work Nb/Re bilayers were grown by molecular beam epitaxy (MBE) on A-plane sapphire, $\alpha$-Al$_2$O$_3$ $(11\bar{2}0)$.  Both Nb (a bcc metal) and Re (hcp) were evaporated via electron beam bombardment at rates of about 0.3~{\AA}/s, with the substrate temperature near 800-850~$^{\circ}$C.  Nb films were typically 1000~{\AA} thick, and the Re thickness was varied from 70-1000~{\AA}.  Films were characterized using \textit{in situ} reflection high-energy electron diffraction (RHEED) and x-ray photo-electron spectroscopy (XPS), and using \textit{ex situ} atomic force microscopy (AFM), x-ray diffraction (XRD), and transmission electron microscopy (TEM).

\begin{figure}[t]
  \includegraphics[width=3.375in]{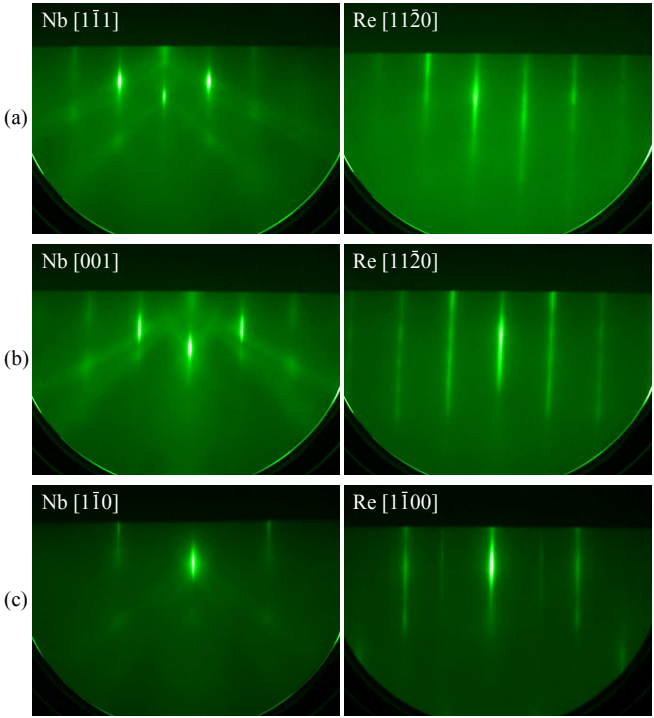}
  \caption{Pairs of Nb (110) and Re (0001) RHEED pictures showing the alignment of the two lattices.  With the RHEED beam parallel to the Nb $[1\bar{1}1]$ axis (a) the Re diffraction image was an asymmetric $[11\bar{2}0]$ pattern, indicating a slight misalignment.  Contrast this with RHEED along the Nb [001] direction (b) where the Re diffraction image was a symmetric $[11\bar{2}0]$ pattern.  Similarly, it was found that the Re $[1\bar{1}00]$ axis was aligned with Nb $[1\bar{1}0]$ (c).  Note that all of the Re RHEED patterns had elongated streaks, and the $[1\bar{1}00]$ image also showed a faint $[11\bar{2}0]$ pattern.  Both observations indicate domain growth with multiple in-plane orientations.}
  \label{rheedpics}
\end{figure}

\begin{figure*}[bt]
  \includegraphics[width=6.500in]{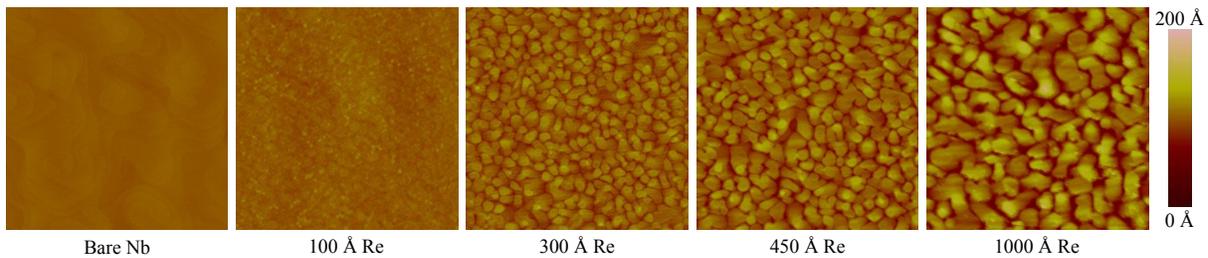}
  \caption{AFM scans covering the full range of Re film thicknesses investigated.  Bare Nb (110) was found to be atomically smooth, while Re (0001) over-layers exhibited evolving surface morphologies characterized by increasing roughness and average grain size.  Scans shown here are all $1\times1$ $\mu$m$^2$ with the height scale shown at right.}
  \label{afmscans}
\end{figure*}

Single-crystal Nb (110) films grown on $\alpha$-Al$_2$O$_3$ $(11\bar{2}0)$ exhibited the orientational relationship
\[\textrm{Nb } [1\bar{1}1] \parallel \alpha\textrm{-Al}_2\textrm{O}_3\textrm{ } [0001] ~,\]
in agreement with the well-established three-dimensional relationship between Nb and sapphire.\cite{Durbin1982,Mayer1990}  Nb RHEED patterns (Fig. \ref{rheedpics}) showed bright streaks and well-defined Kikuchi lines, indicating a high degree of crystallinity.  AFM measurements (Fig. \ref{afmscans}) revealed an atomically smooth surface morphology with an rms roughness typically in the 2-3~{\AA} range.  Sharp Bragg reflections were observed with XRD, and transport measurements commonly showed a residual resistance ratio, $\rho_{293\textrm{K}}/\rho_{10\textrm{K}}$, of about 100.

The Nb (110) surface lattice is very nearly close-packed, and it therefore provides a suitable surface for the heteroepitaxy of close-packed metals.  For Re overlayers the predicted surface plane is (0001), with the in-plane orientation determined by the relative size of the atomic radii and the relative strength of bulk and interface energies.\cite{Bauer1986}  For hcp (0001) metal growth on bcc (110) surfaces, in general there are two in-plane orientations possible.  One is the Nishiyama--Wassermann (NW) relationship,\cite{Wassermann1933,Nishiyama1934} where
\begin{equation}
\textrm{hcp } [11\bar{2}0] \parallel \textrm{bcc } [001] ~. \tag{NW}
\end{equation}
The second is the Kurdjumov--Sachs (KS) relationship,\cite{Kurdjumov1930} where
\begin{equation}
\textrm{hcp } [11\bar{2}0] \parallel \textrm{bcc } [1\bar{1}1] \textrm{ or } [\bar{1}11] ~. \tag{KS}
\end{equation}
A KS-type orientational relationship yields domains in the hcp metal layer that are rotated $10.5^{\circ}$ with respect to each other.

\begin{figure}[t]
  \includegraphics[width=3.375in]{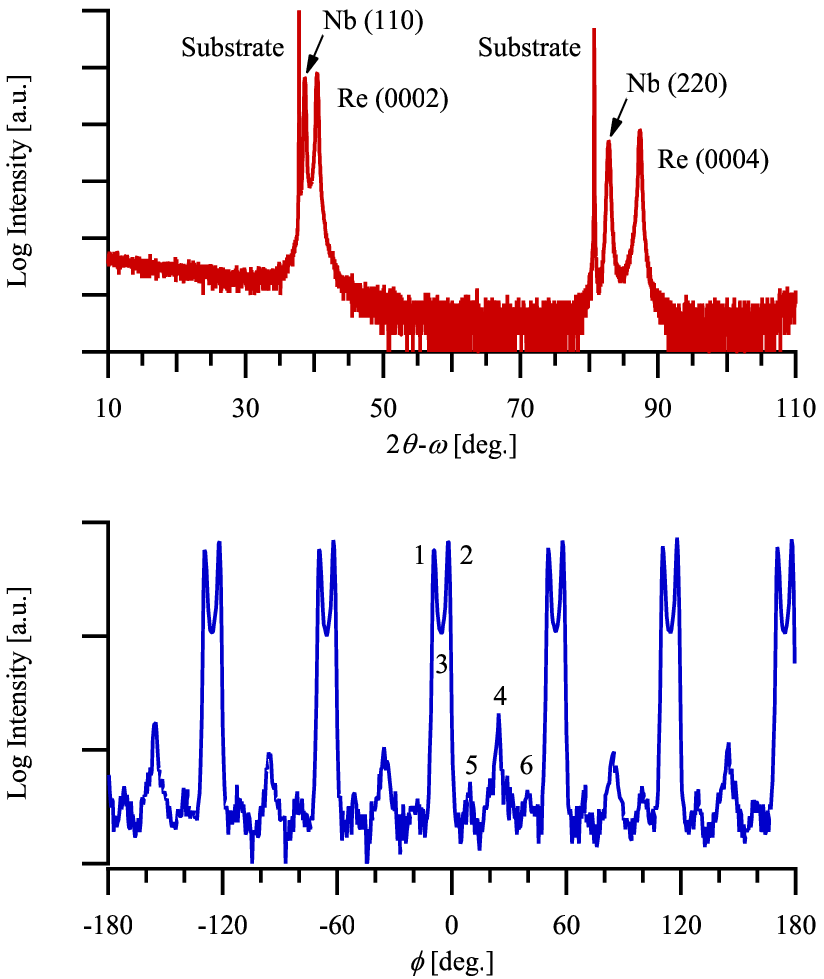}
  \includegraphics[width=3.375in]{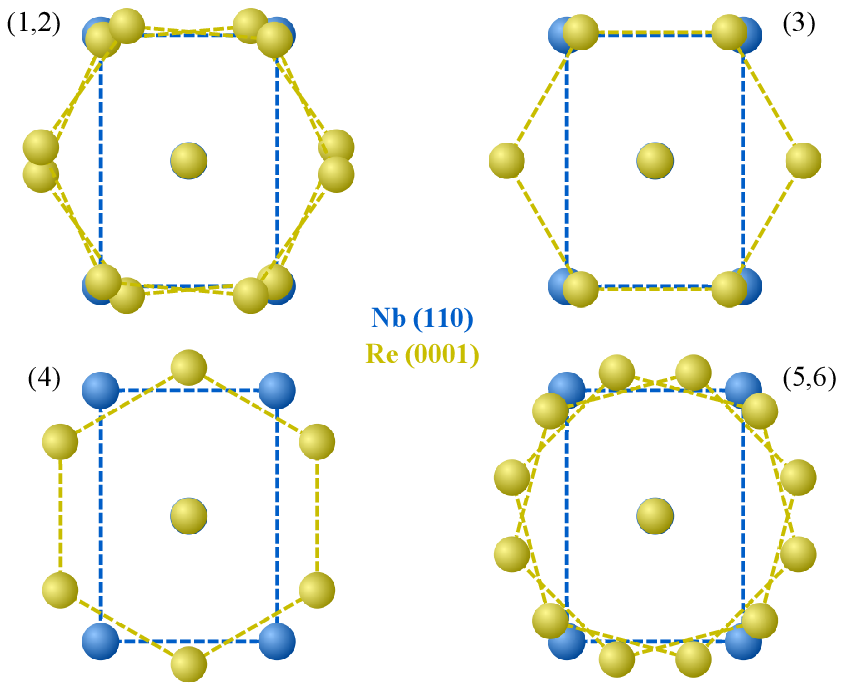}
  \caption{X-ray diffraction scans from Nb/Re bilayers grown on A-plane sapphire.  The $2\theta$-$\omega$ scan exhibits single orientations for both metal layers: (110) for bcc Nb and (0001) for hcp Re.  The $\phi$ scan of off-axis Re $[10\bar{1}1]$ peaks reveals six different in-plane orientations of the Re layer, each of which are depicted here with respect to the underlying Nb (110) surface lattice.  (Numbers on the diagrams correspond to the numbered peaks in the $\phi$ scan.)  98\% of the Re film is comprised of KS- (1,2) and NW-oriented (3) domains.}
  \label{nbrexrd}
\end{figure}

Fig. \ref{rheedpics} shows several RHEED diffraction patterns for both the Nb (110) and Re (0001) surfaces -- the images are paired together to show the alignment of the two layers.  Overall, RHEED images from the Re layer exhibited the hexagonal symmetry that one expects from the (0001) surface.  However, there was ample evidence that multiple in-plane orientations were present, as streaks were vertically elongated and, along some azimuths, the diffraction images had two distinct Re (0001) RHEED patterns superimposed.  With the wafer rotation fixed such that the RHEED beam was along the Nb $[1\bar{1}1]$ azimuth (Fig. \ref{rheedpics}(a)), the corresponding Re diffraction image was an asymmetric $[11\bar{2}0]$ pattern, indicating a misalignment between these two crystal axes.  Contrast this with RHEED along the Nb [001] azimuth (Fig. \ref{rheedpics}(b)), where the Re surface showed a symmetric $[11\bar{2}0]$ diffraction pattern.  Perpendicular to Nb [001] is the $[1\bar{1}0]$ axis (Fig. \ref{rheedpics}(c)), along which the Re diffraction image was a symmetric $[1\bar{1}00]$ RHEED pattern, but also included faint streaks indicative of $[11\bar{2}0]$-oriented domains.  Such RHEED patterns for the Re layer were common for all samples, regardless of Re thickness.

These multiple in-plane orientations are more clearly exhibited in the XRD scans shown in Fig. \ref{nbrexrd}. $2\theta$-$\omega$ scans of all samples showed single surface orientations for both the Nb and Re layers.  However, while the Nb (110) layer grew on A-plane sapphire with a single in-plane orientation, the Re (0001) over-layer exhibited six different in-plane orientations.  Both the KS orientation (peaks labeled 1 and 2 in the $\phi$ scan shown in Fig. \ref{nbrexrd}) and NW orientation (3) were present, and together they comprised about 98\% of the integrated diffraction intensity.  The balance was largely comprised of Re grains oriented with Re $[11\bar{2}0]$ $\parallel$ Nb $[1\bar{1}0]$ (4), consistent with RHEED observations.  Also present -- and representing about 0.2\% of the total signal -- were domains having $[11\bar{2}0]$ $\parallel$ Nb $[1\bar{1}0]$ $\pm 15^{\circ}$ (5 and 6).  The presence and relative abundance of these six in-plane orientations was found to be independent of Re thickness.  Furthermore, for KS-oriented Re grains the angle of separation was $7.3^{\circ}\pm 0.1^{\circ}$ instead of the theoretical value of $10.5^{\circ}$.  All together these findings indicate that the Re overlayer was poorly registered with the underlying Nb (110) surface lattice.

\begin{figure}[b]
  \includegraphics[width=3.375in]{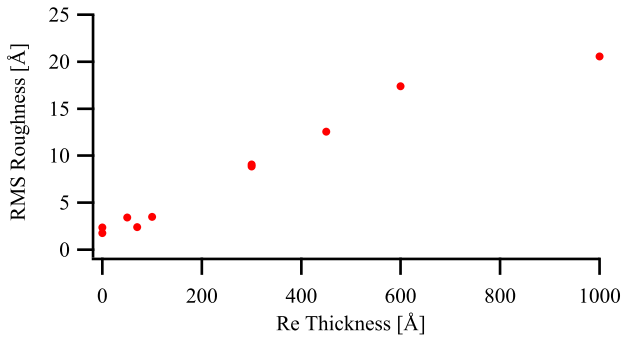}
  \caption{The evolution of Re (0001) surface roughness as a function of film thickness as measured by AFM.  Thin films less than about 150~{\AA} thick retained the atomically smooth surface of the underlying Nb, while thicker films showed very rough surface topologies with large grains.  The scan size used for these measurements was $5\times5$ $\mu$m$^2$.}
  \label{rerough}
\end{figure}

While both the RHEED and XRD analyses showed no dependence on Re thickness, an evolution of the Re (0001) surface morphology was observed.  The heteroepitaxy of Re on Nb (110) best follows the Stranski-Krastanov growth mode,\cite{Bauer1958} whereby the initial stages of growth are layer-by-layer, followed by 3D island growth.  The series of AFM scans shown in Fig. \ref{afmscans} illustrates this well.  Bare Nb (110) surfaces exhibited monolayer step-edges with terrace widths on the order of 1000~{\AA}, and the rms roughness of such surfaces was typically 2-3~{\AA} when measured on a $5\times5$ $\mu$m$^2$ scan.  Thin Re over-layers, less than about 150~{\AA} thick, retained the smooth surface of the underlying Nb to some degree with an rms roughness of less than 5~{\AA} typical.  However, for thicker films the surface roughness increased monotonically up to a Re thickness of 1000~{\AA}, where the rms roughness measured more than 20~{\AA}.  At this thickness the Re (0001) surface morphology was qualitatively very similar to that of Re (0001) films grown directly on C-plane sapphire under similar conditions.  In addition to increased surface roughness, the average grain size observed in surface scans also grew with increasing Re thickness.  The evolution of surface roughness as a function of Re film thickness is plotted in Fig. \ref{rerough}.

\begin{figure}[b]
  \includegraphics[width=3.375in]{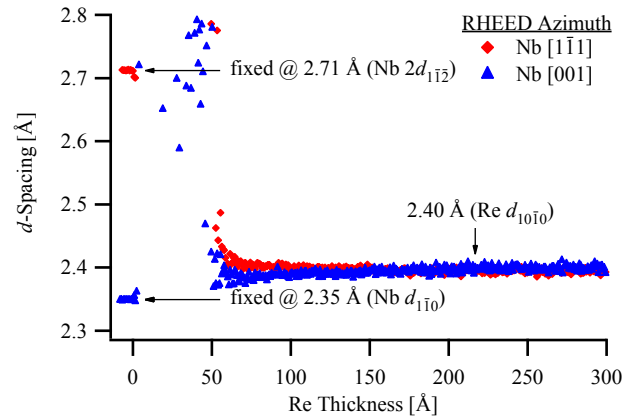}
  \caption{In-plane strain relaxation in Re (0001) films grown on Nb (110) as measured \textit{in situ} with RHEED.  The wafer's azimuthal angle is fixed during Re growth, and assuming an initial surface of bulk-like Nb the lattice constant of the Re layer can be determined.  The measured Re lattice spacing at large thickness agrees well with bulk values, with a fully relaxed lattice by about 200 {\AA}.}
  \label{restrain}
\end{figure}

Strain relaxation in the Re layer was also investigated using RHEED.  Bulk Nb at room temperature has a lattice parameter of 3.300~{\AA},\cite{Barns1968} while Re has an basal plane lattice parameter of 2.761~{\AA}.\cite{Liu1970}  Assuming an NW-type orientational relationship, the lattice misfit between Re (0001) and Nb (110) is -12.9$\%$ and 2.1$\%$ along the Nb $[1\bar{1}\bar{2}]$ and $[1\bar{1}0]$ directions, respectively.  During two separate Re growths the wafer was held fixed with the RHEED beam aligned perpendicular to these directions -- along the $[1\bar{1}1]$ and [001] azimuths, respectively -- giving a direct measure of the surface lattice spacing during growth.  The Re overlayer grew such that the RHEED beam was parallel to (or nearly so) a $[11\bar{2}0]$ axis.  Line scans of the RHEED pattern were digitally captured and analyzed to determine the change in streak spacing as a function of Re thickness.\cite{k-Space}  To obtain a measure of the Re lattice constant, the streak spacing was converted to real-space dimensions by assuming the initial Nb (110) surface was relaxed and bulk-like.  Thermal expansion was also accounted for -- at 825 $^{\circ}$C the fractional change in length with respect to room temperature, $\Delta L/L_{RT}$, for Nb is $6.49\times10^{-3} \pm 3\%$, and for the Re basal plane ($\parallel a\textrm{-axis}$) is $5.65\times10^{-3} \pm 5\%$.\cite{CINDAS}  The resulting strain-thickness curves are shown in Fig. \ref{restrain}.  With the initial $d$-spacing fixed according to the specific RHEED azimuth, the Re layer was found to be completely relaxed by a thickness of 200 {\AA}.  At a thickness of 70 {\AA} the Re layer exhibited an in-plane strain of about 1\%.

\begin{figure}[b]
  \includegraphics[width=3.375in]{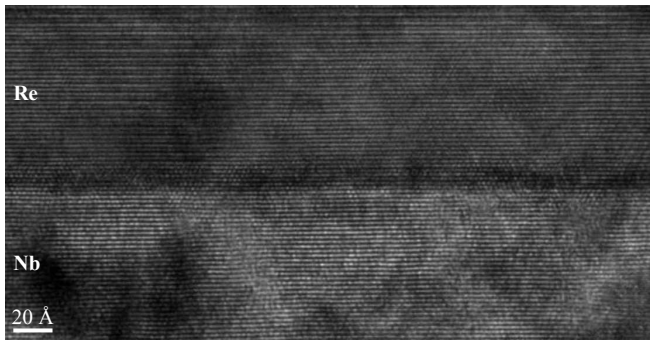}
  \caption{A bright-field TEM image of the Nb/Re interface looking down the Nb $[1\bar{1}\bar{2}]$ axis.  Atomic planes are clearly visible in both layers, but the interface is not atomically sharp.  Instead, there appears to be some mixing over a range of about 5 monolayers, which may responsible for the imperfect registration of the Re (0001) layer with the underlying Nb.}
  \label{renbtem}
\end{figure}

One feature in the strain relaxation curve that stood out was the large number of outliers for Re thicknesses less than about 50~{\AA}.  This was partially due to the fact that measurements were made during Re deposition, when the Re e-gun source was heated to about 2900~$^{\circ}$C and casting light over the entire growth chamber, including the phosphor-coated RHEED screen.  The light from the e-gun caused a substantial increase in the diffraction signal background, so much so that faint RHEED streaks were rendered invisible.  The acquisition and analysis software was not optimized for these conditions, so that RHEED streaks visible to the naked eye at thicknesses as low as 25~{\AA} were not digitally detected.  However, it may also have been the case that the first few Re monolayers did not grow epitaxially -- they may have diffused into or mixed with the underlying Nb.  Evidence of this could be seen in bright-field TEM images, one of which is shown in Fig. \ref{renbtem}.  The interface between the Nb and Re layers was not atomically sharp and instead consisted of a mixed Nb-Re interface region about 5 monolayers thick.  This mixing was not too surprising -- at 800~$^{\circ}$C the solubility limit for Re in Nb is 44 atomic \%, while for Nb in Re it is only about 1 atomic \% \cite{Giessen1961} -- and may be at least partially responsible for the disappearance of RHEED streaks at the onset of Re growth.  It may also explain the poor registration of the Re (0001) layer with the underlying Nb (110) surface.

\section{Discussion}

\begin{figure}[b]
  \includegraphics[width=3.375in]{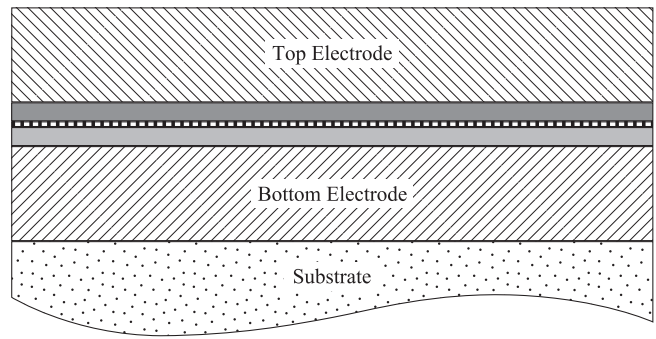}
  \caption{Schematic of a SIS ``trilayer" employing buffer layers between both superconducting electrodes and the insulating tunnel barrier.  While the structure shown here consists of five distinct layers, the structure is designed to behave electrically as a trilayer.  The Nb/Re bilayers described in this Letter may be utilized for the bottom electrode.}
  \label{epi-trilayer}
\end{figure}

The heteroepitaxial growth of atomically smooth Re (0001) layers on Nb (110) surfaces suggests an alternate means to grow all-epitaxial superconductor/insulator/superconductor (SIS) trilayers.  With a Nb/Re bilayer one can take advantage of the superior growth properties of Nb on sapphire (its excellent crystallinity and atomically smooth surface), and the higher Nb critical temperature permits device characterization at 4.2~K. A 100~{\AA} Re overlayer retains the smoothness of the underlying Nb, provides a better lattice match for an epitaxial Al$_2$O$_3$ barrier layer, and resists oxidation better than Nb.\cite{Samsonov1982}  It can also be expected to superconduct well above its natural $T_c$ of 1.7~K due to proximity effects.  In this manner, the Re layer serves as a buffer between the Nb electrode and an oxide tunnel barrier -- it serves as both a diffusion barrier and a structural transition.  It is also conceivable that one would place a buffer layer between the tunnel barrier and top electrode as well.  Such a scheme of epitaxial trilayer growth is shown in Fig. \ref{epi-trilayer} -- while morphologically this structure has five distinct layers, given the appropriate choice of materials and layer thicknesses it should still behave electrically as an SIS trilayer.

The one strike against the use of Nb/Re bilayers for all-epitaxial trilayer growth is the presence of multiple in-plane orientations, as these domains will only serve to promote granular growth modes in all subsequent layers.  Grains in the tunnel barrier in particular could lead to localized gap states,\cite{Choi2009} critical current noise,\cite{vanHarlingen2004} or two-level fluctuators,\cite{Martin2005,Martinis2005} all potential sources of decoherence in quantum bits.  One possible solution to this domain-growth problem is to simply optimize the Re deposition parameters.  Following the 3/8-rule \cite{Flynn1988} for Re would suggest an ideal temperature for epitaxial film growth of about 1025~$^{\circ}$C, about 200~$^{\circ}$C hotter than what has been used in this study.  However, in this temperature realm the Nb (110) surface can be expected to roughen with the formation of facets due to step flow and bunching.\cite{WelanderThesis}  And if mixing is a concern, higher temperatures would likely lead to increased diffusion at the interface.  Changes in deposition rate may also have an impact, though again there is the competition between surface kinetics and bulk diffusion to consider.

Another potential solution to the problem of domains in the Re layer is to simply use another bcc metal in place of Nb.  There are many bcc/hcp (or bcc/fcc) metal pairs for which ideal heteroepitaxial growth cannot be achieved for a variety of reasons -- see, for example, the papers by Ramirez et. al \cite{Ramirez1984} and Bauer et. al \cite{Bauer1986} and references therein.  Nb/Re may simply be one such pair, likely due to the high miscibility of Re in Nb.  In that case, the only solution would be to pick another bcc metal, or perhaps another metal pair altogether, so long as the two metals together satisfy all the requirements for epitaxial SIS multilayers: single-crystal bilayer growth, superconductivity at temperatures reasonable for doing qubit physics, and a surface that is atomically flat, resistive to oxidation, and well lattice-matched to basal plane sapphire.  To the author's knowledge, this work represents the first reported investigation of a bcc/hcp metal pair that satisfies many of these needs.

\section{Summary}

Epitaxial Nb/Re bilayers were grown on sapphire by molecular beam epitaxy.  Thick Nb (110) films were deposited on A-plane sapphire, and thin Re (0001) overlayers were shown to be atomically smooth and relatively strain-free.  Diffraction measurements revealed the presence of multiple in-plane orientations in the Re layer, and TEM images showed evidence of inter-diffusion at the Nb/Re interface.  Despite these shortcomings, this work demonstrates the proposed concept that Nb/Re bilayers (or similar bcc/hcp metal pairs) be utilized for all-epitaxial SIS trilayer growth.  Such an approach may enable the elimination of many materials-induced decoherence mechanisms in quantum information science and technology applications.

\section{Acknowledgements}

The author would like to thank W. D. Oliver and V. Bolkhovsky for fruitful discussions, J. Powers for technical assistance with the MBE system, and D. Calawa for doing the XRD measurements.  This work was sponsored by the Department of the Air Force under Air Force Contract number FA8721-05-C-0002.  Opinions, interpretations, conclusions, and recommendations are those of the author and are not necessarily endorsed by the United States Government.

\bibliography{re_nb_jap}

\end{document}